\pdfoutput=1
\documentclass[submission, PhysProc]{SciPost}

\hypersetup{
    colorlinks,
    linkcolor={red!50!black},
    citecolor={blue!50!black},
    urlcolor={blue!80!black}
}

\DeclareSymbolFont{usualmathcal}{OMS}{cmsy}{m}{n}
\DeclareSymbolFontAlphabet{\mathcal}{usualmathcal}

\newcommand\half{{\frac{1}{2}}}

\graphicspath{{figures/}}

\begin{document}

\begin{center}{\Large \textbf{
Calculation of asymptotic normalization
coefficients in the complex-ranged Gaussian basis\\
}}\end{center}

\begin{center}
D.A. Sailaubek\textsuperscript{1$\star$},
O.A. Rubtsova\textsuperscript{2}
\end{center}

\begin{center}
{\bf 1} Faculty of Physics and Technical Sciences, L.N. Gumilyov Eurasian National University, 010000 Nur-Sultan, Kazakhstan
\\
{\bf 2} Skobeltsyn Institute of Nuclear Physics, Moscow State University, 119991 Moscow, Russia
\\

${}^\star$ {\small \sf s.dinmuhamed@gmail.com}
\end{center}

\begin{center}
\today
\end{center}


\definecolor{palegray}{gray}{0.95}
\begin{center}
\colorbox{palegray}{
  \begin{tabular}{rr}
  \begin{minipage}{0.05\textwidth}
    \includegraphics[width=14mm]{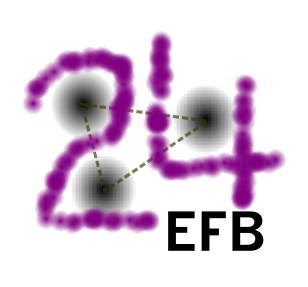}
  \end{minipage}
  &
  \begin{minipage}{0.82\textwidth}
    \begin{center}
    {\it Proceedings for the 24th edition of European Few Body Conference,}\\
    {\it Surrey, UK, 2-6 September 2019} \\
    \end{center}
  \end{minipage}
\end{tabular}
}
\end{center}

\section*{Abstract}
{\bf
A new technique towards finding asymptotic normalization coefficients in the complex-ranged
Gaussian basis is presented. It is shown that a diagonalisation procedure for the total Hamiltonian matrix in the
given basis results in approximation for a radial part of the bound state wave function from the origin
up to the far asymptotic distances, which allows to extract ANCs rather accurately.
 The method is  illustrated by calculations of single-particle ANCs for
nuclei bound states in cases of non-local nucleon-nucleus interactions, in particular, phenomenological global potentials with the Perey-Buck’s non-locality.
}


\section{Introduction}
\label{sec:intro}
The complex-range Gaussian basis (CRGB)
has been demonstrated to give a convenient
representation for bound-state \cite{hiyama2012} and scattering \cite{epja2018} calculations. The distinctive features of the basis
 follow from its definition, i.e. the
basis functions are constructed from the conventional real-valued Gaussians with additional oscillating factors which makes them suitable for approximation of wave
functions of highly excited bound states and
continuum states as well. At the same time, they can be considered as a combination of Gaussians with complex scale parameters
which allows to use all the advantages of the Gaussian expansion approaches.

 Recently, this basis has been
successfully applied for calculation of scattering amplitudes for
charged particles within the Coulomb wave-packet
 formalism \cite{epja2018}. Here we examine the CRGB in
evaluation of asymptotic normalization coefficients (ANCs) which
represent an important information for
a description of nuclear reactions, e.g. for calculation
of the radiative capture cross-sections in nuclear astrophysics.

Below we briefly describe the method of calculation and compare results for single particle ANCs for mirror nuclei
found with local and non-local nucleon-nucleus interactions.

\section{Calculation method}
\subsection{Complex-range Gaussian basis}
The functions of the CRGB are constructed from a set of Gaussians with complex scale parameters:
\begin{equation}\label{eq:psi}
	\psi_{nl}(r)=r^le^{-(1+ib)\alpha_nr^2},\ n=1,2,\dots,K,
\end{equation}
where $b$ and $\alpha_n$ are chosen arbitrarily, in general.
By using a combination of the above complex scale
Gaussian with its complex conjugated, one can introduce the real-valued basis functions in the form:
\begin{equation}\label{eq:phi}
	\phi_{nl}(r)\equiv N_{nl}\frac{1}{2}\left[\psi_{nl}^*(r)+\psi_{nl}(r)\right] = N_{nl}r^le^{-\alpha_nr^2}\cos(b\alpha_nr^2),
\end{equation}
where $N_{nl}$ are normalization factors.

In our practical calculations, we use the fixed value of the parameter $b=\dfrac{\pi}{4}$ and the Tchebyshev grid for the parameters
$\alpha_n$:
\begin{equation}
\label{an}
\alpha_n=\alpha_0\left[\tan\left(\frac{2n-1}{4K}\pi\right)\right]^t,\quad b=1,\ldots,K
\end{equation}
where $\alpha_0$ is the scale parameter and $t$ defines density of the grid. 
The detailed study of the optimal parameter set for the CRGB can be found in our recent paper \cite{epja2018}.
\subsection{Single-particle ANCs for local potentials}
Below we will adopt the above basis for calculation of single particle ANCs for ground states of nuclei.
The nucleon-nucleus potentials are taken in the conventional Woods-Saxon form:
\begin{equation}\label{eq:nucl_pot}
	V_{NA}(r) = V_0f_0(r) + V_{so}(\vec{l}\cdot\vec{s})\frac{\hbar^2}{m_\pi^2c^2}\frac{d}{rdr}f_{so}(r),
\end{equation}
where $f_\alpha(r) = \left[1 + \exp\left(\frac{r-R_\alpha}{a}\right)\right]^{-1}$, $R_\alpha = r_\alpha A^{1/3}$ and
Coulomb interaction has a form of charged sphere at short distances:
\begin{equation}\label{eq:coul_pot}
	V_C(r) = \left\{
	\begin{array}{lc}
	\dfrac{z_1z_2e^2}{2}\left(3 - \dfrac{r^2}{R_C^2}\right) & \text{if } r < R_C,\\
	\dfrac{z_1z_2e^2}{r} & \text{if } r \geq R_C.\\
	\end{array}
	\right.
\end{equation}

ANC is usually defined as the limit of the ratio of the bound state
wave function (the reduced radial part) to the Whittaker function $W_{-\eta_{bs}, l+1/2}(2k_{bs}r)$ \cite{bertulani}:
\begin{equation}\label{eq:ANC}
	u_l(r)\approx b_lW_{-\eta_{bs}, l+1/2}(2k_{bs}r) \quad r\rightarrow\infty, \quad
	k_{bs} = \sqrt{\frac{2m\varepsilon_{bs}}{\hbar^2}},
\end{equation}
where $\eta_{bs}$ is the Coulomb parameter, $k_{bs}$ is the wave number corresponding to the binding energy $\varepsilon_{bs}$ and $m$
 is the reduced mass.

In the approach, the bound state wave function is expanded over the CRGB functions (\ref{eq:phi}):
\begin{equation}
	u_l(r) = r\sum_{n=1}^{K}C_n\phi_{nl}(r).
\end{equation}
Then, the coefficients $C_n$ are found from the generalized eigenvalue
problem for the Hamiltonian $H$ matrix:
\begin{equation}
	\det|H_{nn'} - EI_{nn'}| = 0, \label{hnn}
\end{equation}
where $I_{nn'} = \langle\phi_{nl}|\phi_{n'l}\rangle$ is the overlap matrix. After solving the problem (\ref{hnn}), the
corresponding ANC is found directly from the eq.~(\ref{eq:ANC}) at some distance $r_m$.

\subsection{Account of the Perey-Buck type nonlocality}
The evident advantage of the suggested way of calculation is that the same procedure can be used for non-local form of nucleon-nucleus interaction.
Consider  the same potential as (\ref{eq:nucl_pot}) with the Perey-Buck type non-locality \cite{perey_buck}:
\begin{equation}
\label{nloc}
	U(\textbf{r}, \textbf{r}') = V_0f_0\left(\frac{1}{2}|\textbf{r}+\textbf{r}'|\right)
	\frac{\exp\left[-\left(\dfrac{\textbf{r}-\textbf{r}'}{\beta}\right)^2\right]}{\pi^{\frac{3}{2}}\beta^3}.
\end{equation}
We follow below the usual approximation when the variable $p=\half|{\bf r}+{\bf r}'|$ for the Woods--Saxon shape is replaced with 
variable $\half(r+r')$. In such a case, the usual partial wave expansion is employed and  
the matrix elements of the potential (\ref{nloc}) in the CRGB can be found as integrals:
\begin{equation}
	U_{nn'} = \langle\phi_{nl}|U|\phi_{n'l}\rangle = \int_{0}^{\infty}\int_{0}^{\infty}r^2drdr'\phi_{nl}(r)g_l(r, r')\phi_{n'l},
\end{equation}
where
\begin{equation}
	g_l(r,r') = \frac{2i^lz}{\sqrt{\pi}\beta}j_l(-iz)\exp\left(-\frac{r^2+r'^2}{\beta^2}\right)V_0f_0(p),
\end{equation}
and $z=\frac{2rr'}{\beta^2}$, $p=\frac{r+r'}{2}$.
These matrix elements are calculated numerically.

\section{Results}
\begin{figure}[h]
	\centering
	\includegraphics[width=0.49\textwidth]{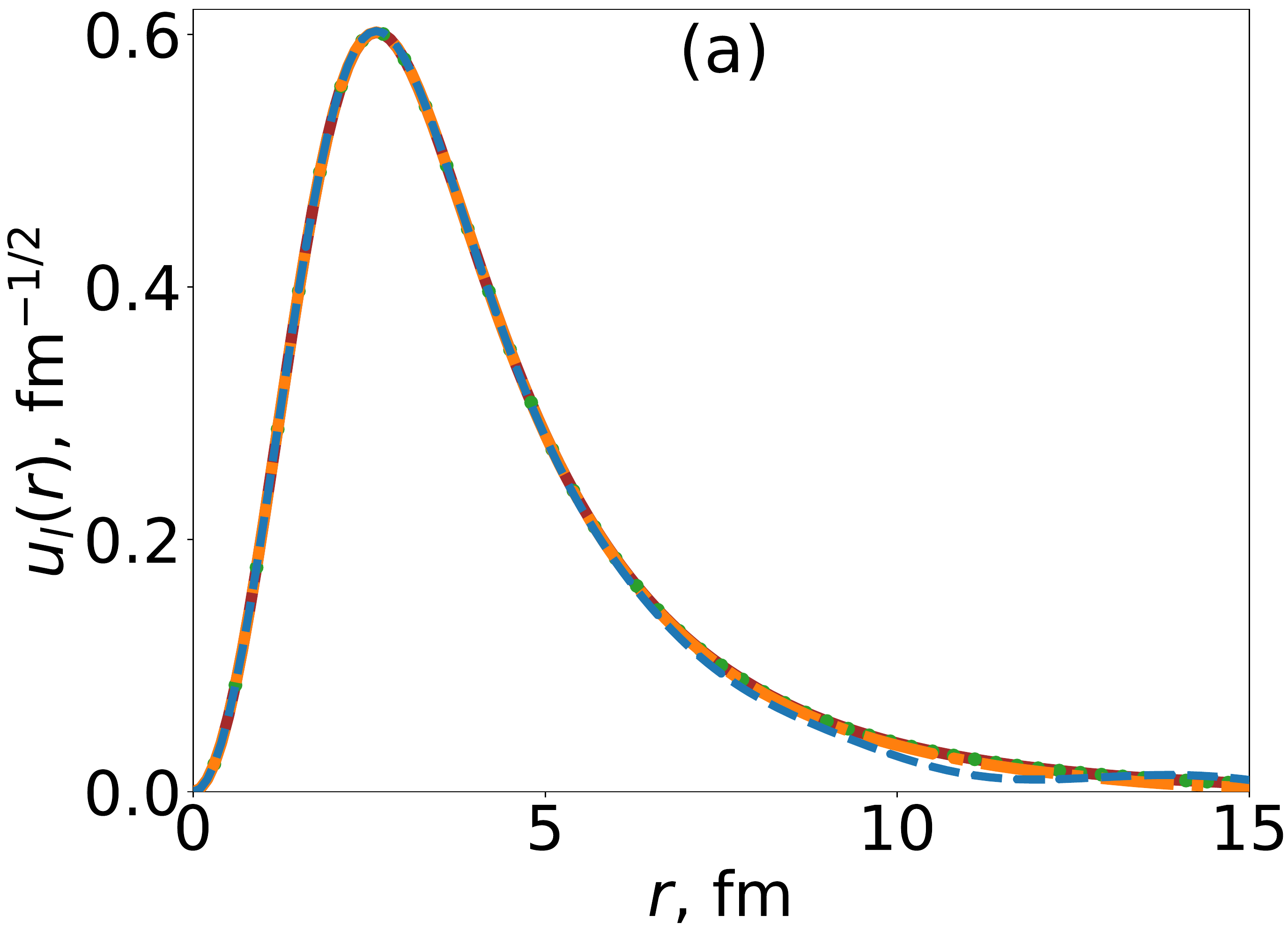}
	\includegraphics[width=0.49\textwidth]{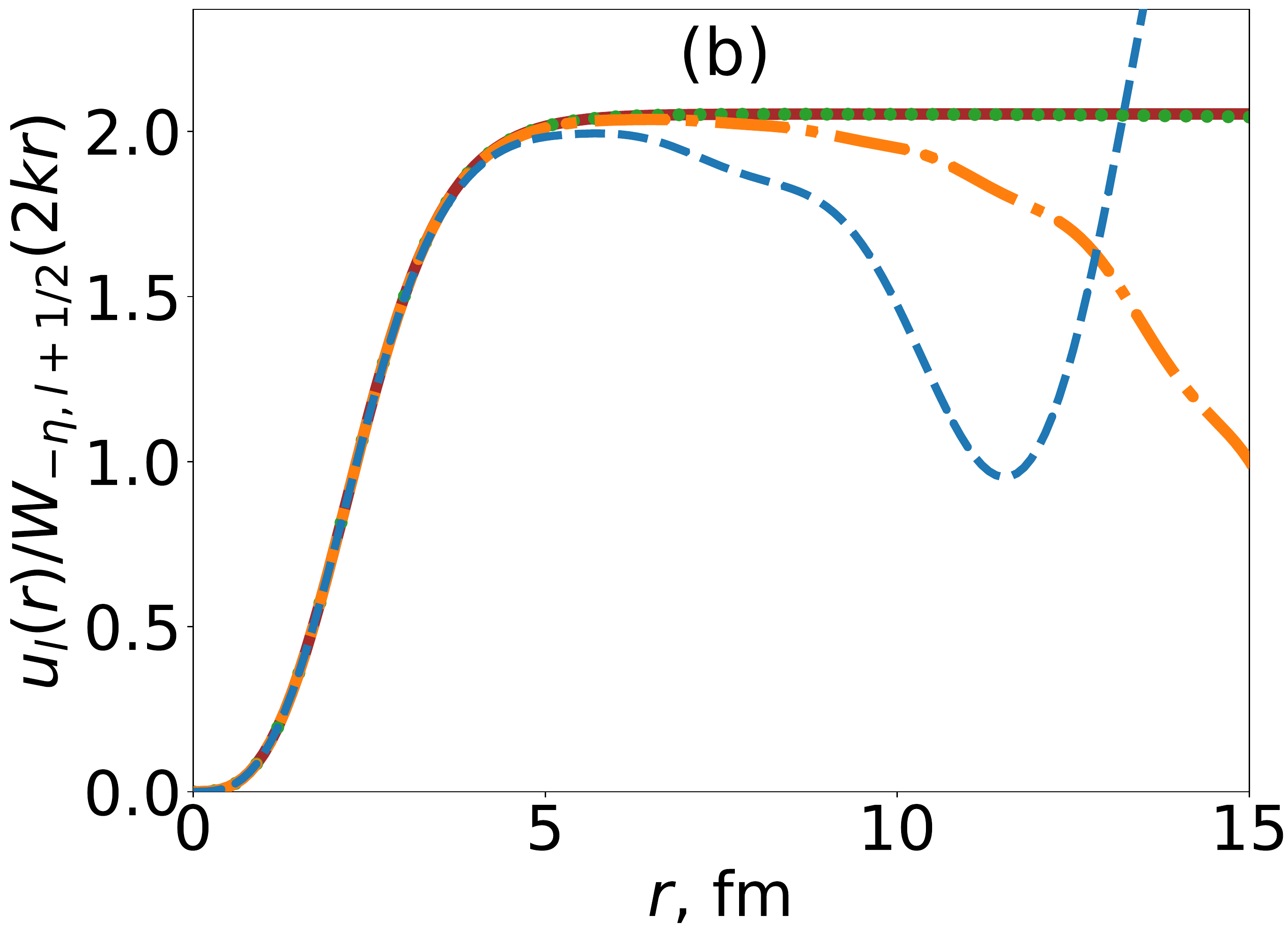}
	\caption{The proton ground state 1$p_{1/2}$ wave function for $p+^{12}$C (a) and its ratio to the Whittaker function (b) obtained using CRGB with $K = 10$ (dashed curves), 20 (dash-dotted curves), 30 (dotted curves), 40 (solid curves).}
	\label{fig:local}
\end{figure}
Consider at first the reaction $^{12}$C(p, $\gamma$)$^{13}$N, where we assume that the proton in the 1$p_{1/2}$ orbital is captured by $^{12}$C to form the ground state of $^{13}$N. In Fig. \ref{fig:local} the wave function of this state and its ratio to the Whittaker function found in CRGB with different basis dimensions $K$ are shown. Here the values $\alpha_0=0.1$ fm$^{-2}$ and $t=3$ for the grid (\ref{an}) have been used. As is seen from  Fig. \ref{fig:local}a, this ratio approaches a constant value at $r_m>5$ fm  with increasing the basis dimension. The basis with $K=30$ is quite enough to get an accurate result.

\begin{figure}[h!]
	\centering
	\includegraphics[width=0.49\textwidth]{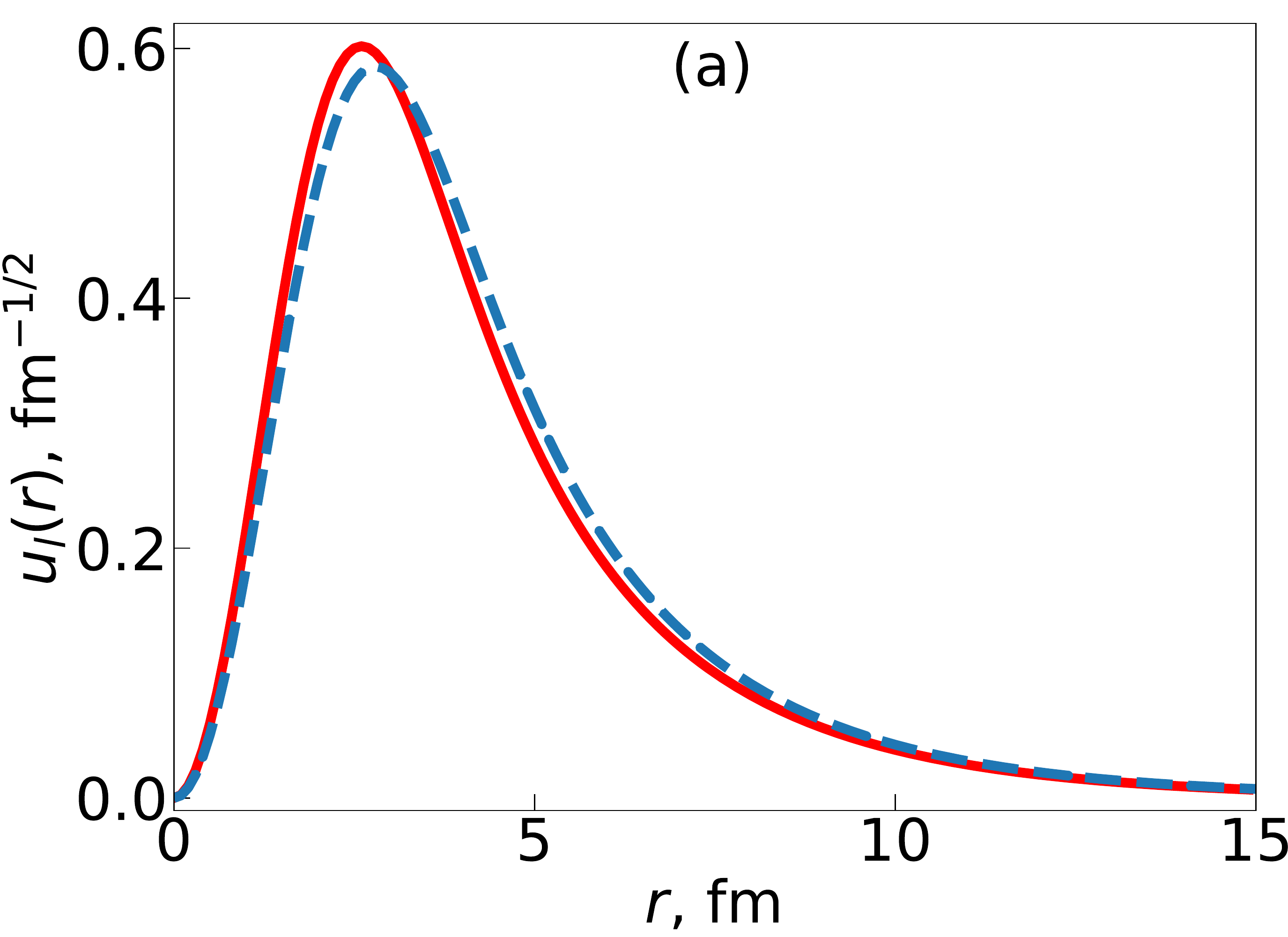}
	\includegraphics[width=0.49\textwidth]{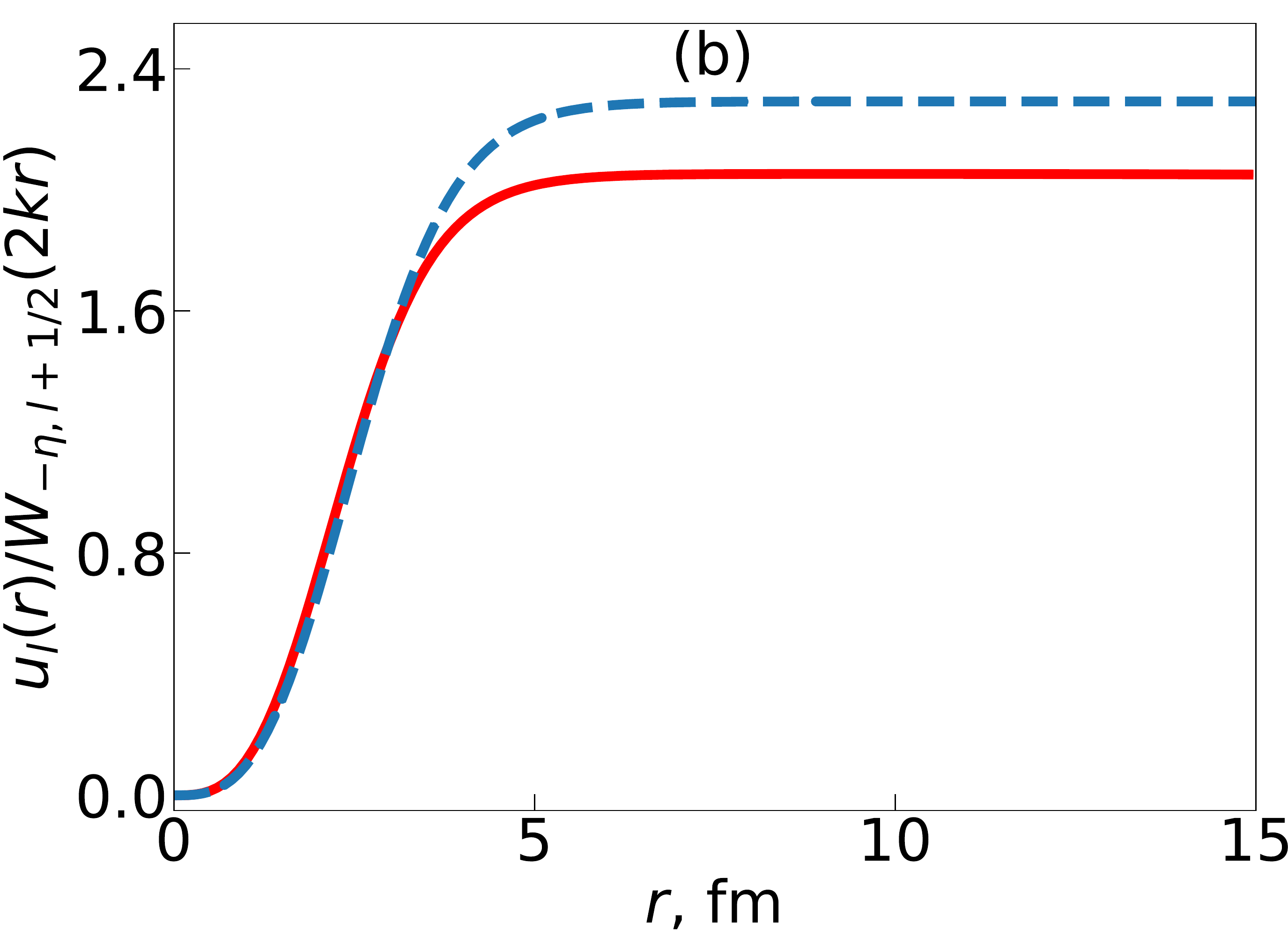}
	
	\caption{The proton ground state 1$p_{1/2}$ wave function for $p+^{12}$C (a) and its ratio to the Whittaker function (b) obtained using local (solid curves) and the Perey-Buck potential (dashed curves).}
	\label{fig:nonlocal}
\end{figure}

In Fig.~\ref{fig:nonlocal}, we present bound state wave functions found with local and non-local interactions for the same system $p+^{12}$C. For this calculation, we use the value of the non-locality parameter $\beta=0.85$ fm. Being compared to the local equivalent case, the nonlocality reduces the amplitude of the wave function  and slightly shifts its  tail  to farer distances in full agreement with  the so-called Perey effect \cite{austern}. Thus, this effect should lead to an increase of the corresponding ANC since the  wave function is always normalized to unity. This result is clearly seen in Fig.~\ref{fig:nonlocal}b.

Finally, we have calculated ANCs for three pairs of mirror nuclei $^8$Li-$^8$B, $^{12}$B-$^{12}$N and $^{13}$C-$^{13}$N by using local and non-local nucleon-nucleus potentials. It should be mentioned that the symmetry relations between ANCs  for such mirror pairs has been established \cite{Timofeyuk} for a case of local interactions.

Table \ref{table:1} summarizes local and non-local potential parameters and single-particle ANCs calculated using CRGB for three pairs of mirror nuclei. The results for local potentials show good agreement with results of other methods \cite{bertulani}. Here we see again that account of non-locality leads to increasing of the corresponding ANCs in comparison with local potential cases.

\begin{table}[h!]
	\caption{Central potential depth ($V_0$, MeV), binding energy ($\varepsilon_{bs}$, MeV) and single-particle ANC ($b_l$, fm$^{-1/2}$) for radiative proton/neutron capture reactions (the unit of $V_{so}$ is MeV, those of $r_0 = r_c = r_{so}$ and $a_0 = a_{so}$ are fm). Local potential parameters are taken from ref.~\cite{bertulani}.}
	\begin{center}
		\begin{tabular}{lp{1.8cm}p{1.2cm}p{0.7cm}p{0.8cm}p{0.8cm}p{0.8cm}l}
			\hline
			\hline
			\rule[.8ex]{0pt}{1.5ex} Reaction & Group & $V_0$ & $V_{so}$ & $r_0$ & $a_0$ & $\varepsilon_{bs}$ & $b_l$ \\
			\hline
			\rule[.8ex]{0pt}{1.5ex} $^7$Be(p, $\gamma$)$^{8}$B & local & -41.26 & -9.8 & 1.25 & 0.52 & 0.15 & 0.70 \\
			\rule[.8ex]{0pt}{1.5ex} & nonlocal & -45.80 & -9.8 & 1.25 & 0.52 & 0.14 & 0.77 \\
			\rule[.8ex]{0pt}{1.5ex} $^7$Li(n, $\gamma$)$^{8}$Li & local & -43.56 & -10 & 1.195 & 0.65 & 2.03 & 0.76 \\
			\rule[.8ex]{0pt}{1.5ex} & nonlocal & -48.21 & -10 & 1.195 & 0.65 & 2.03 & 0.82 \\
			\hline
			\rule[.8ex]{0pt}{1.5ex} $^{11}$C(p, $\gamma$)$^{12}$N & local & -40.72 & -10 & 1.25 & 0.65 & 0.63 & 1.50 \\
			\rule[.8ex]{0pt}{1.5ex} & nonlocal & -44.29 & -10 & 1.25 & 0.65 & 0.60 & 1.63 \\
			\rule[.8ex]{0pt}{1.5ex} $^{11}$B(n, $\gamma$)$^{12}$B & local & -34.33 & -10 & 1.25 & 0.65 & 3.39 & 1.35 \\
			\rule[.8ex]{0pt}{1.5ex} & nonlocal & -37.42 & -10 & 1.25 & 0.65 & 3.37 & 1.46 \\
			\hline
			\rule[.8ex]{0pt}{1.5ex} $^{12}$C(p, $\gamma$)$^{13}$N & local & -41.65 & -10 & 1.25 & 0.65 & 1.94 & 2.05 \\
			\rule[.8ex]{0pt}{1.5ex} & nonlocal & -45.48 & -10 & 1.25 & 0.65 & 1.94 & 2.28 \\
			\rule[.8ex]{0pt}{1.5ex} $^{12}$C(n, $\gamma$)$^{13}$C & local & -41.35 & -7 & 1.236 & 0.62 & 4.95 & 1.85 \\
			\rule[.8ex]{0pt}{1.5ex} & nonlocal & -45.46 & -7 & 1.236 & 0.62 & 4.95 & 2.08 \\
			\hline
			\hline
		\end{tabular}
	\end{center}
	\label{table:1}
\end{table}

\section{Conclusion}
In this study we have shown that the expansion of bound state wave functions in the CRGB allows to reproduce their asymptotics rather accurately including cases when long-range Coulomb interaction is taken into account.

The suggested method for calculation of ANCs treats non-local interactions without any additional difficulty.

The obtained ANCs in a case of non-local nucleon-nucleus interactions occur to be greater than those for the local potentials for the same systems. This result is fully consistent with the Perey effect \cite{austern} and previous calculations for non-local potentials (see e.g. \cite{tian2018}).

The method developed can be generalized directly to calculation of ANCs for few-body systems by using 
the few-body complex ranged Gaussian basis.

\section*{Acknowledgements}
The authors thank Dr. N.K. Timofeyuk for useful comments and suggestions. O.A.R.
appreciates the partial financial support from the RFBR grant 19-02-00014.



\bibliography{Sailaubek1}

\nolinenumbers

\end{document}